# Naming game with learning errors in communications

**Yang LOU** and **Guanrong CHEN***

Department of Electronic Engineering, City University of Hong Kong, Hong Kong SAR, China

*Corresponding author: eegchen@cityu.edu.hk

Abstract

Naming game simulates the process of naming an objective by a population of agents organized in a certain communication network topology. By pair-wise iterative interactions, the population reaches a consensus state asymptotically. In this paper, we study naming game with communication errors during pair-wise conversations, where errors are represented by error rates in a uniform probability distribution. First, a model of naming game with learning errors in communications (NGLE) is proposed. Then, a strategy for agents to prevent learning errors is suggested. To that end, three typical topologies of communication networks, namely *random-graph*, *small-world* and *scale-free* networks with different parameters, are employed to investigate the effects of various learning errors. Simulation results on these models show that 1) learning errors slightly affect the convergence speed but distinctively increase the requirement for memory of each agent during lexicon propagation; 2) the maximum number of different words held by the whole population increases linearly as the value of the error rate increases; 3) without applying any strategy to eliminate learning errors, there is a threshold value of the learning errors which impairs the convergence. The new findings help to better understand the role of learning errors in naming game as well as human language development from a network science perspective.

## 1. Introduction

Naming game (NG) is a simulation-based experimental study that explores the emergence of shared lexicons in a communicating population of agents about a same object which they observed. In the minimal version of naming game [1], agents reach a consensus state after individual iterative *actions* and pair-wised local *interactions* among them, where *action* means learning words from external lexicons or creating words in case one has nothing in its memory and *interaction* include propagation of words among agents as well as checking the state of consensus. By testing on different models under various conditions, such as networks with realistic topologies and functional parameters as well as broadcasting and learning strategies *etc*., it is possible to reveal some features and even principles of linguistic conventions via such extensive computer simulations.

In a general NG model, a population of agents with either finite or infinite memory is employed [2], where the agents may or may not have words in their memories initially [3]. The rule of the game is as follows: At each iteration, a speaker and a hearer are picked from the population at random, and the speaker transmits one word either from its memory to the hearer, or if the speaker has nothing in its memory then it picks up a word from a large enough dictionary (external lexicon), which is equivalent to creating a new word by itself. If coincidently the hearer has the same word as the one that the speaker named, then they have consent and consequently they both clear their memories except keeping the common word; otherwise, the hearer adds the new word to its own memory as a result of learning this word from the speaker. Through running such an iterative interaction and propagation process until the game converges to a steady state where every agent keeps one and only one common word in its memory.

It is worth mentioning that the speaker and hearer must be neighbors to be able to communicate to each other, which is reflected by the connectivity of the population network in a certain topology.

Recently, Li *et al.* [4] studied the following scenario of the NG: At each iteration, one speaker with multiple hearers are picked simultaneously from the population at random, and then the game rule is followed throughout. This model is referred to as naming game with multiple hearers (NGMH). The simulation results show that multiple hearers accelerate the convergence speed, as compared to the original NG model. Based on the NGMH mode, Gao *et al*. [5] further investigated a generalized version with multiple speakers and multiple hearers (called naming game in groups, or NGG), in which it lets every agent in a selected group from the population to play the role of both a speaker and a hearer simultaneously. It demonstrated that the convergence speed is faster when the number of participating agents increases.

Considering the real-life scenario of human communications, the efficiency of agent learning and information propagation affect the convergence of a naming process. Language acquisition is generally error-prone. But, interestingly, as pointed out by Nowak *et al*. [6], learning errors can help prevent the linguistic system from being trapped in sub-optimum situations by increasing the diversity, thereby leading to the evolution of a more efficient language where the system is evaluated by the function of payoff. Nowak *et al*. [6] also found some thresholds of the error rate for certain learning models, below which the system gains advantage from learning errors while above which mistakes will impair the system, say, reducing the payoff. Moreover, *noise* may lead to recurrently converging states of a Markov chain model [7], which is considered beneficial in better detecting social interactions. Therefore, errors or noises may be expected to affect the language system positively, to some extent, in the two NG models introduced above. However, our study in this paper includes neither the formation of a more efficient language as in [6], nor to obtain a series of recurrently converging states as in [7]. Instead, we study learning errors from the perspective of language communications in a common sense that learning errors likely bring negative effects. Every individual makes mistakes sometimes in real life, but if an individual is more experienced then he will probably make less mistakes or even know how to avoid making mistakes. From this consideration, we study a real-life scenario where agents are all error-prone initially but then they gradually learn to avoid making further errors therefore eventually all agents are error-free in the NG process, so that the whole population will reach consensus asymptotically.

Specifically, we will study NG with communication errors during pair-wise conversations, where errors are represented by error rates in a uniform probability distribution. An NG model with learning errors in communications (NGLE) is first proposed, followed by a strategy for agents to prevent learning errors, in Section 2. Three typical topologies of communication networks (*random-graph*, *small-world* and *scale-free* networks) with different parameters are investigated in Section 3, to reveal the effects of various learning errors on the convergence performance, with simulation results presented and analyzed. Finally, conclusions are drawn in the last section.

## 2. Naming game with learning errors in communications

### 2.1. Learning errors in communications

Errors in communications could be from different sources such as the speakers, the communication media, and the hearers. There are also many reasons that could lead to errors, for example ambiguous pronunciations, different styles of handwriting, and imperfect coding and decoding techniques of telecommunications, *etc*., as shown in Fig. 1 where all the boxes and arrows in the figure could lead to errors.

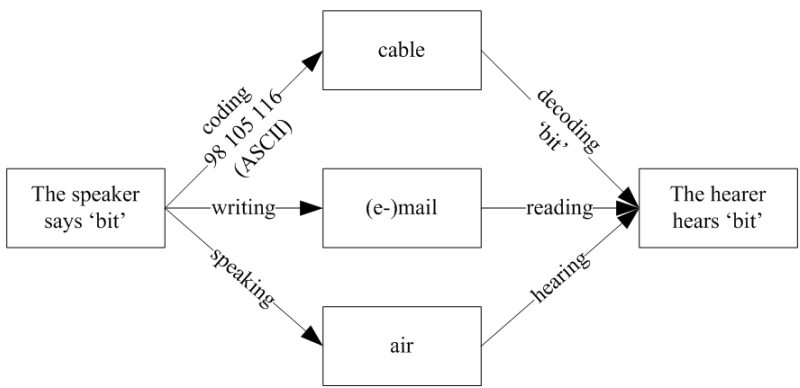

Fig. 1 Methods of communication between a speaker and a hearer

In this paper, we do not study the reason and source of communication errors, but investigate how learning errors propagate and affect the NG convergence. We assume that the occurrence of any type of error can be represented by a numerical value, *error rate*. We also assume that the results caused by different types of error are equivalent to the situation that *the hearer learns a wrong word from the speaker*, which is represented by *learning error* and measured by *error rate*. For example, a word *'bit'* sent by the speaker will be learned by the hearer correctly under a certain probability, while under its complementary probability the hearer will receive a wrong word. Fig. 2 provides an illustration, where the conditional judgment rhombus represents all the different communication media shown in Fig. 1.

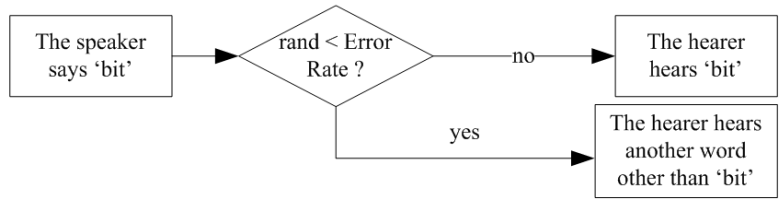

Fig. 2 The communication model with learning errors

In real life, both people and machines produce errors, but it is not acceptable if it happens perpetually, since people will learn and machines can be modified to work correctly. The NG system includes multiple agents and a great deal of lexicons, where lexicons are objective and would not be wrong themselves, but un-experienced agents are error-prone. In an NG, the agents should be able to learn to prevent or at least reduce learning errors by themselves, or they would be considered unreliable. There are many ways for the agents to be more reliable. Empirically and practically, training or educating could be one effective way to improve the correctness of agents' expressions and understandings in their conversations. Double-checking before sending out a word and after receiving it is another solution. In addition, using a redundant medium or some redundant information, *e.g.*, sending out an email accompanied with an auxiliary voice message, or just attaching a redundant message to verify if the hearer receives the information timely and correctly.

In the proposed NGLE model, we adopt the first method mentioned above. We assume that, when an agent has experience as a speaker before, it will have lower probability to make mistakes. This follows the old saying that *teaching and learning grow hand in hand*, which implies that, when an agent has played the role as a speaker, it benefits from the experience, so it is an experienced agent and would rarely make mistakes. As a result, in the early stage of lexicon propagation, communications among the agents are mainly error-prone, while in the later stage the probability of error appearing would be gradually decreased until all agents have participated as speaker at least once, such that the population will finally converge.

*2.2. The NGLE model*

In the NGLE model, an agent in the population is represented by a node in the network. If two agents are neighbors, the corresponding two nodes in the network are connected by an edge; therefore, they can communicate with each other. Thus, the terms *node* and *agent* are interchangeable throughout the paper.

The network model of NGLE is summarized as follows.

1. A population of *n* agents connected in a certain topology is initialed with empty memories, along with an external vocabulary of very large size.
2. At each iteration, a speaker is randomly picked from the population:
    2.1. If the speaker has empty memory, then it randomly picks a *word* from the external vocabulary;
    2.2. Otherwise, the speaker randomly picks a *word* from its memory.
3. A hearer is randomly picked from the neighborhood of the speaker:
    3.1. If the hearer has never been a speaker before, then within the error rate $\rho$, it receives *word'* other than *word*;
    3.2. Otherwise, the hearer receives the *word* correctly.
4. The hearer checks if it has the same word in memory as the received one:
    4.1. If the *word* was already in its memory, learning is successful, so both the speaker and the hearer clear out their memories except keeping the only *word* which was just communicated;
    4.2. Otherwise, the hearer adds the new *word* into its memory.
5. Repeat Step 2 to Step 4 iteratively, until all nodes keep one and only one same word, or until the number of iterations reaches a pre-defined (large enough) value of termination.

Step 3 is illustrated in Sec. 2.1, where all types of possible errors are presented by a single numerical value, the *error rate*. In the following section, both Step 3 and Step 4, namely the process of communication, is interpreted. It is quite different from the minimal NG [1], which has two possible results (*success* and *failure*) after one iteration in communication, in that the NGLE has four possible situations.

## 2.3. The communication process

At each iteration step, a speaker is randomly selected from the population, and so is a hearer but from the speaker's neighborhood represented by the connected edges in the network. This is a direct strategy [8, 9] of agent selection, in which hub nodes have a lower probability to be selected as a speaker, as compared with the so-called reverse strategy [8], which selects the hearer first and then selects a speaker from its neighbors, both at random.

The communication process starts as soon as both the speaker and the hearer have been selected. As shown in Fig. 3(1), the first situation is that there is no learning error, thus the word *signal* is sent from the speaker to the hearer directly and correctly. Then, the hearer checks the received word with its memory and finds *signal* is not included therein, so the hearer adds the word *signal* into its memory. This situation is named "failure without learning error"; it is analogous to the situation that the hearer directly learns a new word *signal* from the speaker. The second situation shown in Fig. 3(2) is the state of consensus, where the hearer has the word *bite* sent by the speaker and both of them erase their memories but keep the common word *bite* only. These two situations are exactly the same as that in the minimal naming game [1].

Now, there are two more new situations in the NGLE. The third and fourth situations are related to learning errors. In Fig. 3(3), the speaker says a word *right* to the hearer, under a certain probability (error rate), the hearer receives a wrong word *night*; after checking its memory, the hearer learns a "new word" that was not included in its memory. Note that if the speaker says a word that should lead to consensus (namely, the hearer actually had it in memory), then in this case they both miss an opportunity to be successful.

The fourth situation is an interesting one, where the speaker says *right* but the hearer receives a wrong word *light*, and coincidently the hearer holds the word *light*. Then, an ambiguous consensus happens: from the consensus reaction of the hearer, the speaker considers that the hearer agrees with his word *right*, while the hearer actually agrees with *light*. This situation is called pseudo consensus. The result of pseudo consensus is that the speaker deletes all words from its memory other than *right*, while the hearer clears out it memory leaving only *light*. This is analogous to the misunderstanding between the speaker and the hearer in real life, while neither of them realizes the existence of an error.

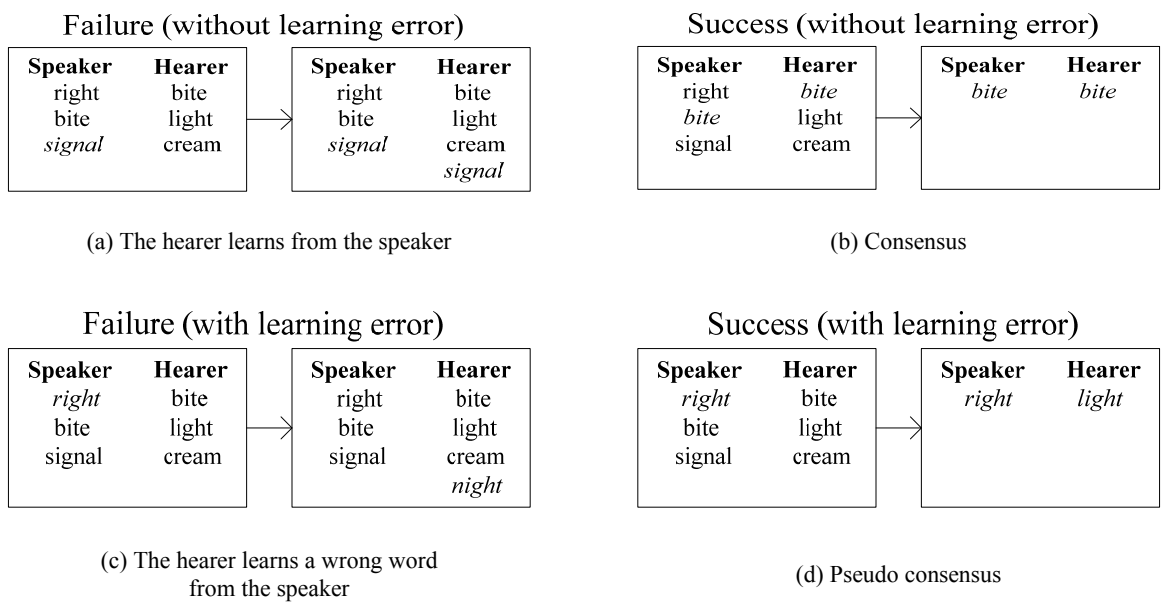

Fig. 3  Four situations in one iteration during pair-wise communication

## 3. Simulated experiments and discussion

The detailed processes of emergence, propagation and consensus of shared lexicons are examined by employing three typical network topologies, *random-graph* (*RG*) [10], *small-world* (*SW*) [11] and *scale-free* (*SF*) networks [12]. As suggested by Baronchelli *et al*. [13], we perform extensive numerical simulations to study the various naming games with comparisons. As implemented in [4], when the memory of the picked speaker is empty, it will randomly pick a word from the vocabulary instead of randomly creating a new word; similarly, when learning error occurs, the hearer will randomly pick a word from the vocabulary and this word should be different from the word received.

### 3.1. Simulation setup

We simulate nine types of networks, which each network contains 2,000 nodes. To reduce the randomness and increase the confidence level, for each type of network, we perform 20 independent simulations and then take an average. Hence, the data shown in Table 1 and the curves shown from Fig. 4 to Fig. 7, are all statistically averaged results from the 20 independent simulation trials. The size of the external vocabulary is 10,000.

Table 1 shows the detailed network setting. We employ three types of *random-graph* networks by altering the connection probability, including *RG*/0.03, *RG*/0.05 and *RG*/0.1, and three types of *scale-free* networks by altering the number of adding nodes at each step, including *SF*/25, *SF*/50 and *SF*/75. For *small-world* networks, they all have two parameters to adjust, *i.e.*, the number of neighborhoods and the rewiring probability, thus six combinations are obtained by altering each parameter with three values each, including *SW*/20/{0.1, 0.2, 0.3} and *SW*/40/{0.1, 0.2, 0.3}. For brevity, the parameter setting of SW/40/{0.1, 0.2, 0.3} is shown in Table S1 of the Supplementary Information (SI) [14], and the simulation results are presented and analyzed in the rest of SI.

Table 1 Network settings in simulations

| Notation | Network Type | Number of nodes | Average degree | Average path length | Average clustering coefficient |
|---|---|---|---|---|---|
| *RG*/0.03 | Random-graph network with *P* = 0.03 | 2,000 | 59.97 | 2.1305 | 0.0300 |
| *RG* /0.05 | Random-graph network with *P* = 0.05 | 2,000 | 99.96 | 1.9564 | 0.0500 |
| *RG* /0.1 | Random-graph network with *P* = 0.1 | 2,000 | 199.92 | 1.9000 | 0.1000 |
| *SW*/20/0.1 | Small-world network with *K* = 20 and *RP* = 0.1 | 2,000 | 40.00 | 2.8251 | 0.5360 |
| *SW* /20/0.2 | Small-world network with *K* = 20 and *RP* = 0.2 | 2,000 | 40.00 | 2.6963 | 0.3806 |
| *SW* /20/0.3 | Small-world network with *K* = 20 and *RP* = 0.3 | 2,000 | 40.00 | 2.6133 | 0.2597 |
| *SF*/25 | Scale-free with 26 initial nodes and 25 new edges added at each step | 2,000 | 49.66 | 2.2312 | 0.0760 |
| *SF* /50 | Scale-free with 51 initial nodes and 50 new edges added at each step | 2,000 | 98.69 | 1.9725 | 0.1217 |
| *SF* /75 | Scale-free with 76 initial nodes and 75 new edges added at each step | 2,000 | 147.10 | 1.9273 | 0.1602 |

Because the different initial states, say one-word-per-agent or no words at all, they would generate different convergence curves [3]. For the purpose of studying the consensus patterns of the NGLE, we assume that there is no initial word in the memory of each agent [4]. We study comprehensively on the values of the learning error rate, $\rho$, which varies from 0.001 to 0.009 by an increment of 0.001, and from 0.01 to 0.09 by an increment of 0.01, and from 0.1 to 0.5 by an increment of 0.1, respectively, together with the reference group having $\rho = 0$. Thus, totally 24 groups are studied. We did not consider the situation with error rate greater than 0.5, since it means most of the information propagated is incorrect, which is out of question.

The last parameter to introduce is the maximum number of iterations, which is set to be 10,000,000 in our simulations. This value is empirically large enough for the nine networks with 24 error rates. Therefore, in each single run, the population definitely reaches a consensus state before the termination value of ten million iterations.

*3.2.    Convergence process*

First, we analyze the relationship between the *number of total words* in the population and different values of the error rare. For clarity, we list only 5 representative sets of data out of 24 sets in total. Some common grounds for all types of networks include that the *number of total words* starts from zero, since we assume that agents are memory-free initially, and then it goes through an ascending (learning is dominant, Figs. 3(a) and (c)) and descending process (consensus is dominant, Figs. 3(b) and (d)), and finally it converges to 2,000 words, which is exactly the number of agents. In the consensus state, each agent holds one and only one same word.

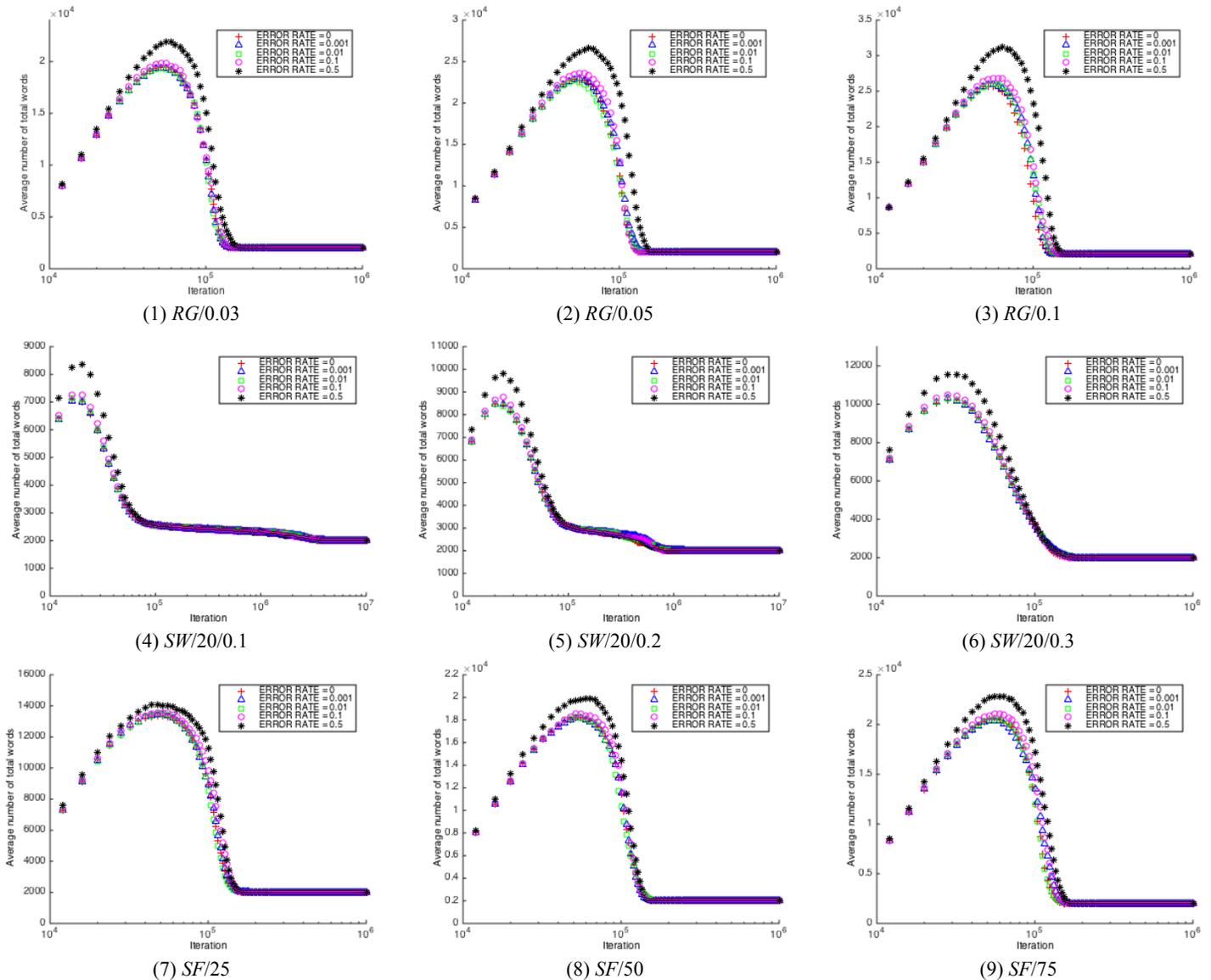

(1) *RG*/0.03    (2) *RG*/0.05    (3) *RG*/0.1
(4) *SW*/20/0.1    (5) *SW*/20/0.2    (6) *SW*/20/0.3
(7) *SF*/25    (8) *SF*/50    (9) *SF*/75

Fig. 4  The convergence process in terms of the number of total words in a network

As can be seen from Fig. 4, two types of networks (Figs. 4 (4) *SW*/20/0.1, and (5) *SW*/20/0.2) converge after more than 1,000,000 iterations, while the rest seven converge in less numbers of steps. The convergence curves of the error rate less than or equal to 0.01 cannot be visually distinguished from the curves without learning errors. But when the error rate increases to 0.1, the curve difference is recognizable, and when the error rate is 0.5, the curve is significantly different from other curves. This means that when the error rate is less than or equal to 0.01, the influence of the learning error on the number of total words is insignificant, while if the error rate is 0.1, it becomes non-negligible, and for 0.5, it becomes quite significant, which means that it generates more words for agents to store in their memories temporarily (but they will be dropped finally). Therefore, from the viewpoint of memory cost, when the error rate is less than 0.01, it does not require more memory than that without learning errors; however, when it increases to 0.1, the extra memory cost is recognizable, and when it is 0.5, the cost is quite significant.

Next, we analyze the relationship between the *number of different words* and different values of the error rare during the convergence process. The *number of different words* displaces a similar convergence curve as that of the *number of total words*, which starts with zero and then converges to one in the consensus state. This means that finally the whole population holds one same word only, but it is impossible to predict which word it will converge to. As can be seen from Fig. 5, for all nine networks, when the error rate is less than or equal to 0.01, the difference to the non-error curve is negligible, but when the error rate is 0.1, the curve is almost as twice higher as those with the error rate less than or equal to 0.01 during a long period before settling to consensus. Fig 5 actually tells the same story as Fig. 4 does; that is, an error rate less than or equal to 0.01 brings an insignificant influence to the required size of memory, but error rate 0.1 cannot be ignored, while 0.5 is significant.

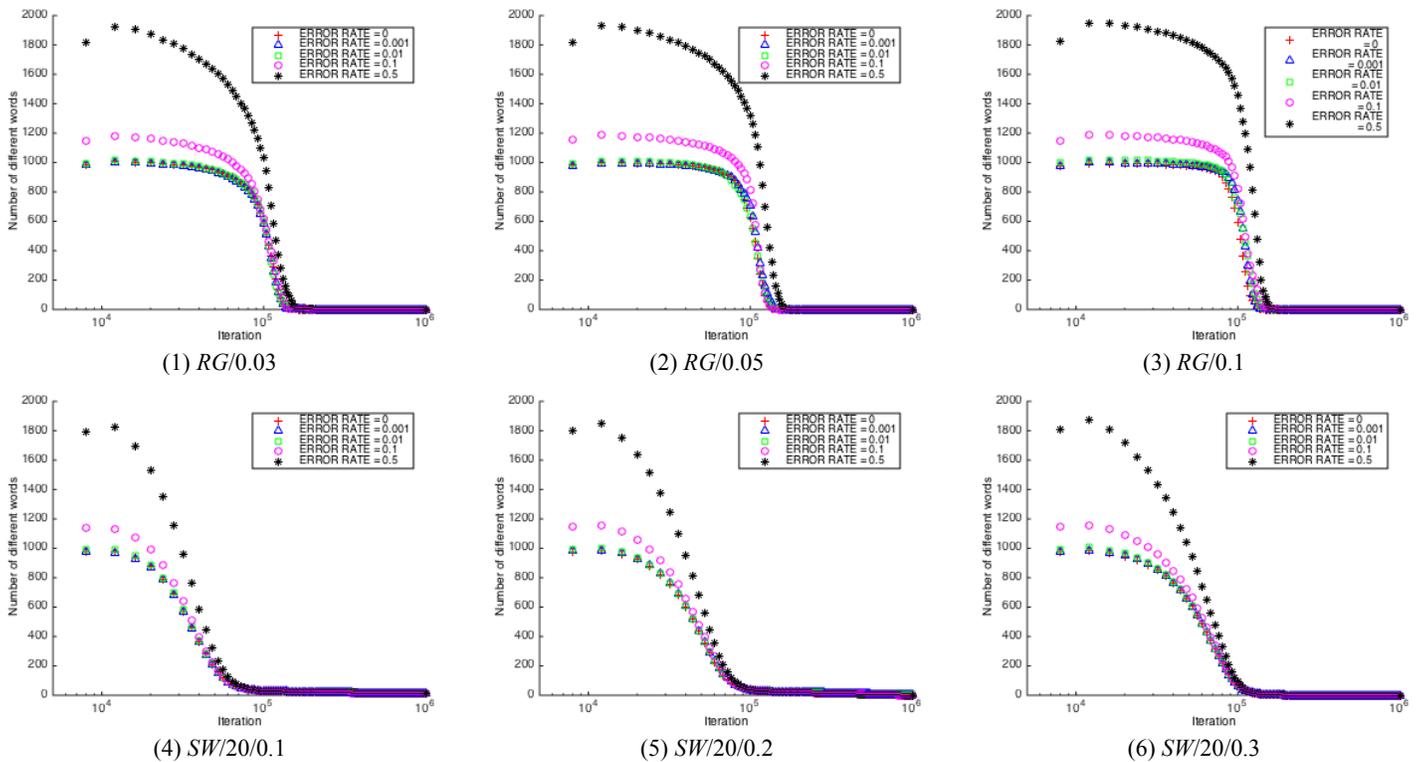

(1) *RG*/0.03  (2) *RG*/0.05  (3) *RG*/0.1
(4) *SW*/20/0.1  (5) *SW*/20/0.2  (6) *SW*/20/0.3

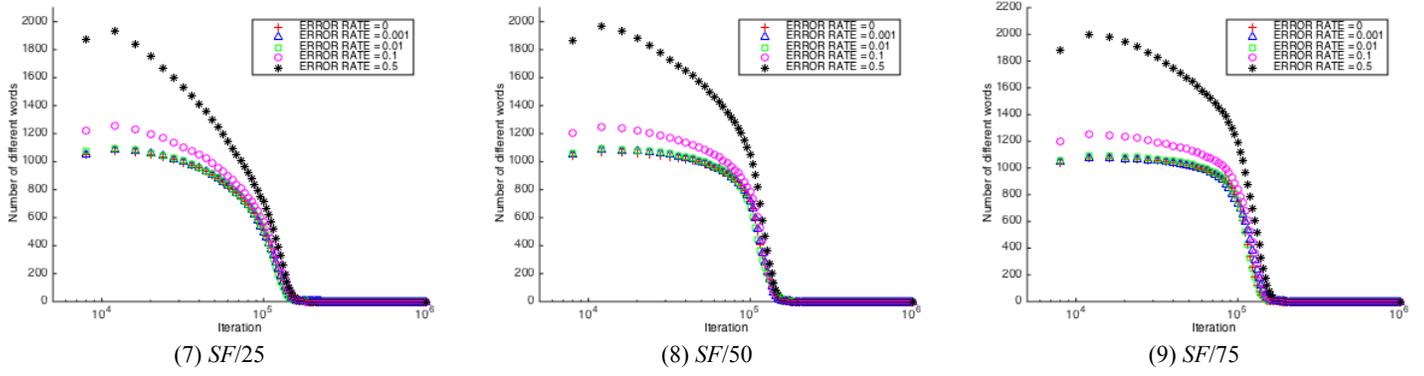

| | (7) *SF*/25 | (8) *SF*/50 | (9) *SF*/75 |

Fig. 5  The convergence process in terms of the number of different words in a network

One explanation for the phenomena shown in Figs. 4 and 5 is that a larger learning error rate gives more chance to randomly pick a word from the vocabulary, so that it is more probable to bring new and different words into the population when learning errors exist, therefore the number of total words and the number of different words both increase when the learning error rate is high.

*3.3.  Convergence time and success rate*

In Section 3.1, we found that when influenced by learning errors, it may require more memory for storage. Now, we examine the convergence time, or the number of iterations, when all agents reach consensus.

Table 2 The increment relationship between convergence time and different values of the error rate

| Networks / Error Rate | *RG*/0.03 | *RG*/0.05 | *RG*/0.10 | *SW*/20/0.1 | *SW*/20/0.2 | *SW*/20/0.3 | *SF*/25 | *SF*/50 | *SF*/75 |
|---|---|---|---|---|---|---|---|---|---|
| 0 | 1.0000 | 1.0000 | 1.0000 | 1.0000 | 1.0000 | 1.0000 | 1.0000 | 1.0000 | 1.0000 |
| 0.001 | +0.0116 | +0.0490 | +0.0768 | +0.0115 | +0.0935 | +0.0275 | +0.0232 | +0.0155 | +0.0431 |
| 0.002 | −0.0154 | +0.0770 | +0.0092 | +0.0412 | −0.0392 | −0.0206 | +0.0412 | +0.0143 | +0.0098 |
| 0.003 | +0.0113 | +0.0455 | **+0.0828** | −0.0514 | −0.0955 | +0.0143 | +0.0169 | −0.0428 | +0.0230 |
| 0.004 | −0.0121 | +0.0635 | +0.0751 | +0.0279 | −0.1055 | +0.0496 | +0.0236 | −0.0121 | +0.0627 |
| 0.005 | −0.0213 | +0.0790 | +0.0653 | +0.0399 | **−0.1581** | +0.0144 | +0.0389 | +0.0802 | +0.0241 |
| 0.006 | −0.0175 | +0.0146 | +0.0898 | +0.0001 | −0.0008 | +0.0062 | +0.0648 | +0.0062 | +0.0735 |
| 0.007 | −0.0502 | +0.0712 | +0.0788 | **−0.0967** | −0.0935 | **+0.0666** | −0.0187 | +0.0486 | +0.0279 |
| 0.008 | −0.0609 | +0.0300 | +0.0579 | +0.0242 | −0.0551 | +0.0377 | +0.0397 | +0.0165 | −0.0072 |
| 0.009 | −0.0166 | −0.0475 | +0.0962 | −0.0297 | −0.0537 | +0.0017 | +0.0064 | −0.0079 | +0.0707 |
| 0.01 | +0.0193 | +0.0165 | +0.1122 | +0.0483 | −0.0040 | +0.0168 | −0.0088 | −0.0228 | +0.0290 |
| 0.02 | −0.0301 | +0.0408 | +0.1176 | −0.0325 | −0.0668 | +0.0352 | +0.0385 | −0.0129 | +0.0915 |
| 0.03 | +0.0255 | +0.0192 | +0.0863 | +0.0911 | −0.0595 | **−0.0513** | −0.0062 | +0.0114 | +0.0121 |
| 0.04 | +0.0152 | +0.0267 | +0.0979 | +0.1052 | −0.1037 | +0.0385 | +0.0575 | +0.0560 | −0.0030 |
| 0.05 | −0.0356 | +0.0470 | +0.1305 | **+0.1472** | +0.0461 | +0.0419 | +0.0024 | −0.0019 | +0.0146 |
| 0.06 | +0.0425 | +0.0718 | +0.0917 | −0.0162 | +0.0484 | +0.0297 | +0.0094 | +0.0266 | **+0.0557** |
| 0.07 | −0.0267 | +0.0320 | +0.0964 | −0.0155 | −0.1183 | +0.0592 | −0.0051 | −0.0064 | +0.0638 |
| 0.08 | +0.0264 | +0.0249 | +0.1304 | −0.0280 | −0.0843 | +0.0090 | −0.0050 | −0.0268 | +0.0500 |
| 0.09 | +0.0193 | −0.0007 | +0.0816 | −0.0799 | **+0.1211** | +0.0359 | **+0.0795** | **+0.0406** | +0.0271 |
| 0.1 | +0.0360 | +0.0205 | +0.1087 | +0.0260 | +0.0262 | −0.0078 | +0.0556 | +0.0279 | +0.0689 |
| 0.2 | +0.0241 | **+0.0734** | +0.2003 | −0.0034 | +0.0769 | +0.0383 | +0.0263 | +0.0260 | +0.0720 |
| 0.3 | +0.0383 | +0.1368 | +0.1524 | +0.1204 | +0.0074 | −0.0193 | +0.0617 | +0.0887 | +0.0154 |
| 0.4 | **+0.0854** | +0.1609 | +0.1622 | +0.0430 | −0.0764 | +0.0449 | +0.0332 | +0.0701 | +0.1406 |
| 0.5 | +0.0886 | +0.1719 | +0.2103 | −0.0334 | −0.0604 | +0.0009 | +0.0473 | +0.0321 | +0.1265 |
| Summary | 13+/10− | 21+/2− | 23+/0− | 13+/10− | 7+/16− | 19+/4− | 18+/5− | 15+/8− | 19+/2− |

Table 3 Statistical division of the numbers of different incremental values of the data in Table 2 into different intervals

| Increment value intervals | $(-\infty, -0.2)$ | $[-0.2, -0.1)$ | $[-0.1, 0)$ | $[0, 0.1)$ | $[0.1, 0.2)$ | $[0.2, +\infty)$ |
|---|---|---|---|---|---|---|
| Number of data | 0 | 4 | **53** | **132** | 16 | 2 |

Table 2 provides the accurate information that cannot be seen directly from Fig. 4. In Table 2, we set the convergence time for the case without learning errors as the benchmark, which is 1.0000 in the first row of the table, and in the following rows, the increment values of convergence time affected by different learning errors are shown. For example, +0.0116 (in row 2, column 1) means that, in the network with *RG*/0.03, when the error rate is 0.001, its convergence time increases by 0.0116 compared to that without learning errors, while with error rate 0.002, it requires only 0.9846 convergence time (−0.0154 in row 3, column 1) of that without learning errors.

In the cases of random-graph and scale-free networks, there exist some approximate thresholds, when error rates are greater than that the increments of convergence time become positive and relatively large. For the random-graph networks, the threshold of RG/0.03 is 0.4, for RG/0.05, it is 0.2, and for RG/0.1, it is 0.003. For the three scale-free networks simulated, they are 0.09, 0.09 and 0.06, respectively. In contrast, such threshold does not appear in small-world networks. There is not an observable relationship between error rates and increments of the convergence time. We denote the maximum positive/negative increments in bold. For the cases of networks *SW*/20/0.1 and SW/20/0.3, the learning errors generally increase the convergence time, while for *SW*/20/0.2, they generally reduce it.

The last row of Table 2 summarizes the number of cases that increase and reduce the convergence time, respectively. For the cases of networks *SW*/20/0.2, learning errors speed up the convergence faster than the postponement, while in the rest six networks, learning errors generally delay the convergence.

In general, a relatively large error rate, above 0.06, will delay visibly the convergence of both random-graph and scale-free networks. But when the error rate is smaller than or equal to 0.05, whether the influence is delaying or accelerating is uncertain. In the case of small-world networks, whether the learning errors speed up or delay the convergence depend on the parameters setting of a network. For *SW*/20/0.2, the error rates mostly speed up the convergence, while for *SW*/20/0.1 and *SW*/20/0.3, they mostly delay the convergence. However, the value of delay or acceleration in the same small-world network is also uncertain but small.

Table 3 gives the statistical numbers of different increment values, which reports that the most error-caused convergence delay are located within [0, 0.1), and the second is [−0.1, 0). Statistically, learning errors bring small increments to the convergence time.

The curves of success rates shown in Fig. 6 are consistent with the information given by Table 2. We use the term *success rate* in a similar fashion as defined in [4], but consider also the pseudo consensus situation as a success. Thereby, the success rate is calculated by the number of iterations of (pseudo) consensus during each of 10 iterations, divided by 10. In Figs. 6(2), it reflects the success rates of small-world networks, the difference between success rate curves is not clearly distinguishable; however, in Figs. 6(1) and (3), they are somewhat more observable. The success rate curves of the other networks, including the rest six networks listed in Table 1 and small-world SW/40/{0.1, 0.2, 0.3}, are shown in SI.

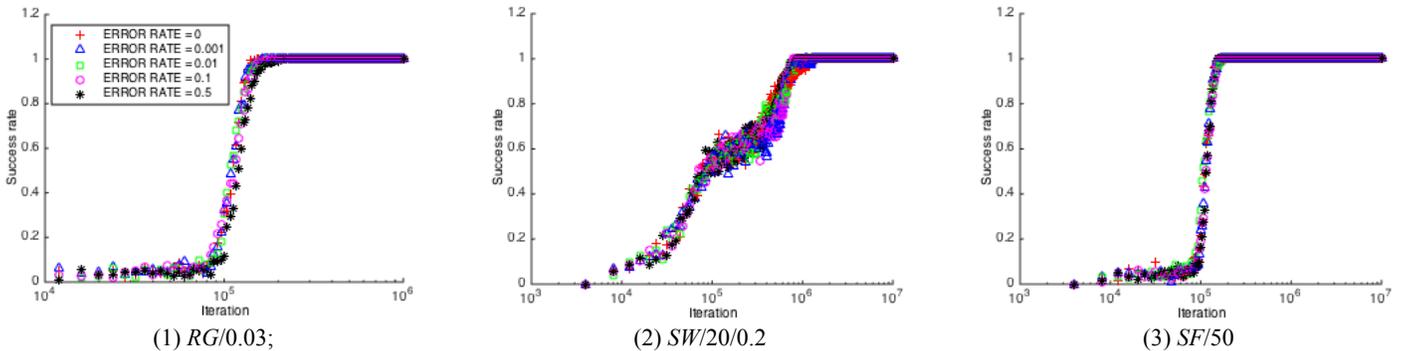

(1) *RG*/0.03;  (2) *SW*/20/0.2  (3) *SF*/50
Fig. 6 Success rate curves

## 3.4. Maximum number of different words

As learning errors affect the required memory size more than the convergence delay, we next examine the relationship between the *maximum number of different words* and different error rates.

In the NGLE model, we assume that 1) agents are possible to have learning errors, and 2) agents will learn to avoid making further errors by being speakers. Figs. 7(1), (2), and (3) show three random-graph networks; three small-world networks, where each agent has 20 neighbors and 40 neighbors, respectively, and three scale-free networks. The figures in this case show an almost perfectly linear relationship between the *maximum number of different words* and the error rate.

In addition, we repeat the simulations under the same conditions, except for the second assumption above, thus agents make errors throughout the process. In this case, the population may not converge at all, therefore we stop the simulation after 10,000,000 iterations. With 20 independent-runs, the averaged results are shown in Figs. 7(4), (5), and (6). These four figures show mainly three differences compared with the figures lying on the left. First, without any strategy to avoid continuously making errors, the magnitude of the maximum number of words (*i.e.*, the magnitude of vertical coordinates) becomes much larger than that on the left; second, the figures on the right show some quadratic features, while that on the left are relatively straight; and finally, on the left the relationship curves of the same network with different parameters are very close to each other, while on the right they diverge from each other with large gaps.

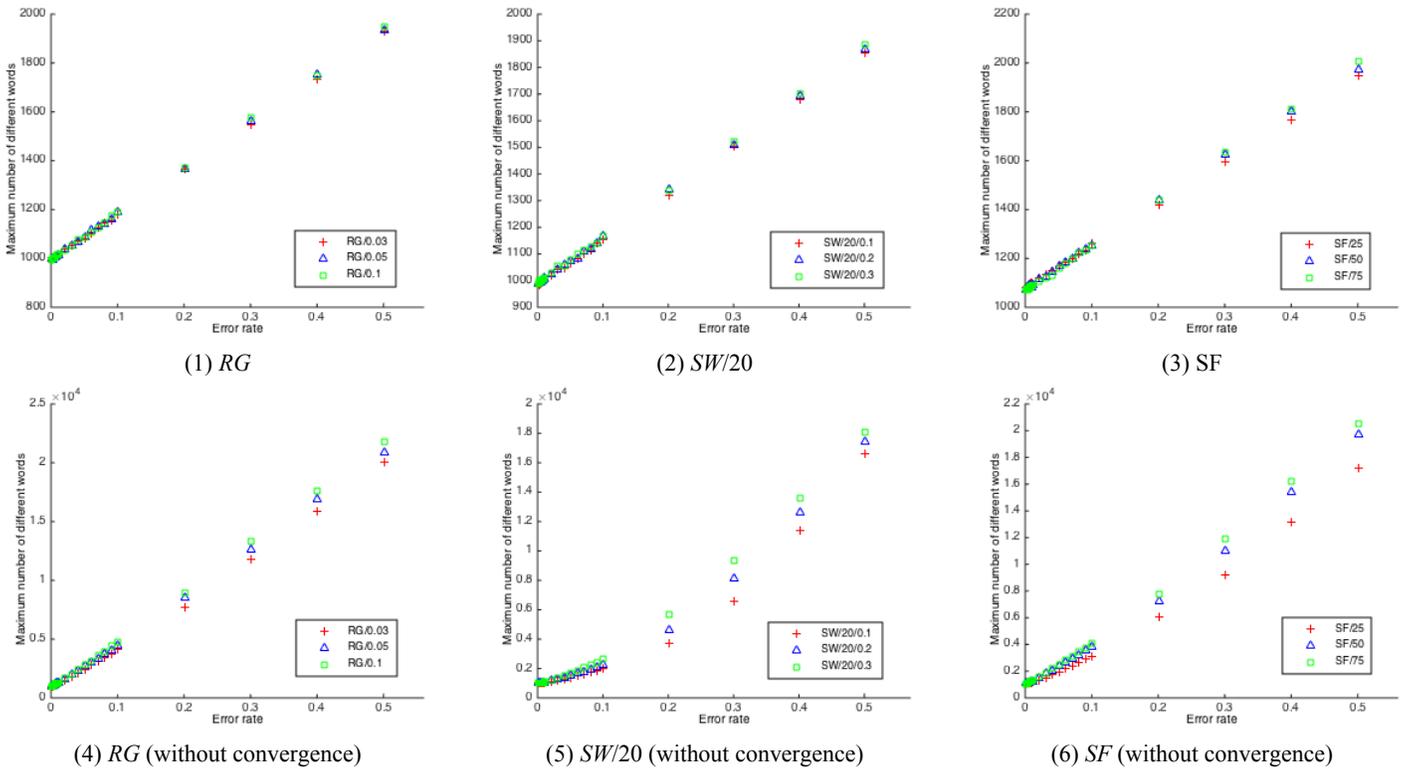

(1) *RG*     (2) *SW*/20     (3) SF

(4) *RG* (without convergence)     (5) *SW*/20 (without convergence)     (6) *SF* (without convergence)

Fig. 7 The relationship between the error rate and the maximum number of different words

## 3.5. Convergence thresholds

To ensure convergence, namely for the population of agents to reach a consensus state before the number of iterations exceeds 10,000,000, the NGLE model employs a rule that when an agent has once been a speaker, it would not make learning error anymore in the future. In the case that all agents continuously make errors during communications, the population may not converge within the pre-defined maximal number of iterations, or even never converge.

As proposed by Nowak *et al*. [6], there is an optimal error rate that maximizes the performance of the parental learning and role model learning system, but for the random learning model this would not work at all when error rate is greater than a small threshold. Since we do not use the system payoff to evaluate the performance, we study the threshold of the error rate that affects the convergence of the population.

Fig. 8 shows the statistic results of the simulations on the threshold. For each type of network, we carry out 20 independent simulations. For each single simulation, we set the initial error rate be 0. If, under the current error rate, the population converges within the maximal number of iterations, then we increase the error rate by an incremental step size of 0.0001. This process repeats until the population does not converge under a certain error rate, and then we record the rate as the threshold of this simulation. It is worth mentioning that, not only suggested by [6], but also by our trial-and-error simulations, incremental step of error rate less than 0.0001 makes no sense to the results for networks having less than 2,000 nodes.

For each box figure shown in Fig. 8, the blue box represents that the central 50% data lie in this section; the red bar is the median value of all 20 datasets; the upper and lower black bars are the greatest and least values, excluding outliers; and finally the red pluses represent the outliers. As can be seen from the box figure, all the random-graph and small-world networks have similar thresholds for the error rate that locate between 0.0061 and 0.0067. More specifically, in most cases, when the error rate increases to around 0.0067 or above, the population would probably not converge within the maximal number of iterations. For the three scale-free networks, this threshold is around 0.0068 to 0.0073, which means that the tolerance of learning errors in scale-free networks is higher than in random-graph and small-world networks.

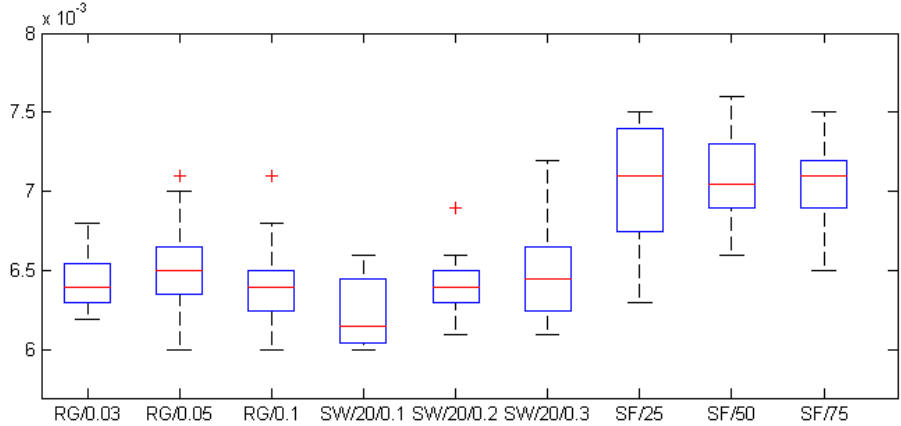

Fig. 8  The box plot of learning error rate threshold in different types of networks

## 4. Conclusions

In this paper, we proposed a novel model of naming game with learning errors in communications (NGLE) and study it by means of extensive and comprehensive simulation experiments. We found that if the agents have some learning errors but can learn to avoid making further errors, the convergence will be slightly affected, either accelerated or delayed within a relative small range, which depend also on the different topologies and parameter settings of the underlying communication networks. However, during the period between the initial and the final consensus states, agents in the NGLE learn and discard more words than the agents without learning errors, which means that learning error requires more memory space from the agents. We also realized that the NGLE model has an interestingly linear relationship between the maximum number of different words throughout the convergence process and the error rate, *i.e.*, the higher the error rate is, the larger the maximum number of different words it will have, which are in a linearly proportional relation. In addition, we identified the statistical range of the error rate threshold, above which the population would not converge in the case without any strategy to prevent learning errors.

It is believed that the new findings reported in this paper are meaningful and helpful to enhance our understanding of the role of learning errors in naming games as well as in evolution of languages.

Acknowledgement

This research was supported by the Hong Kong Research Grants Council under the GRF Grant CityU1120/14.

# Supplementary information for the manuscript
## "Naming game with learning errors in communications"

**Yang LOU** and **Guanrong CHEN*** 
Department of Electronic Engineering, City University of Hong Kong, Hong Kong SAR, China
*Corresponding author: eegchen@cityu.edu.hk

The networks of small-world $SW/40/\{0.1, 0.2, 0.3\}$ are investigated in the following. The parameter settings are shown in Table S1, followed by the simulation results and statistical analysis.

Table S1    Small-word network with number of neighborhoods $K = 40$ with different rewiring probabilities

| Notation | Network Type | Number of nodes | Average degree | Average path length | Average clustering coefficient |
|---|---|---|---|---|---|
| $SW/40/0.1$ | Small-world network with $K = 40$ and $RP = 0.1$ | 2,000 | 80.00 | 2.4499 | 0.5457 |
| $SW/40/0.2$ | Small-world network with $K = 40$ and $RP = 0.2$ | 2,000 | 80.00 | 2.2367 | 0.3894 |
| $SW/40/0.3$ | Small-world network with $K = 40$ and $RP = 0.3$ | 2,000 | 80.00 | 2.1291 | 0.2718 |

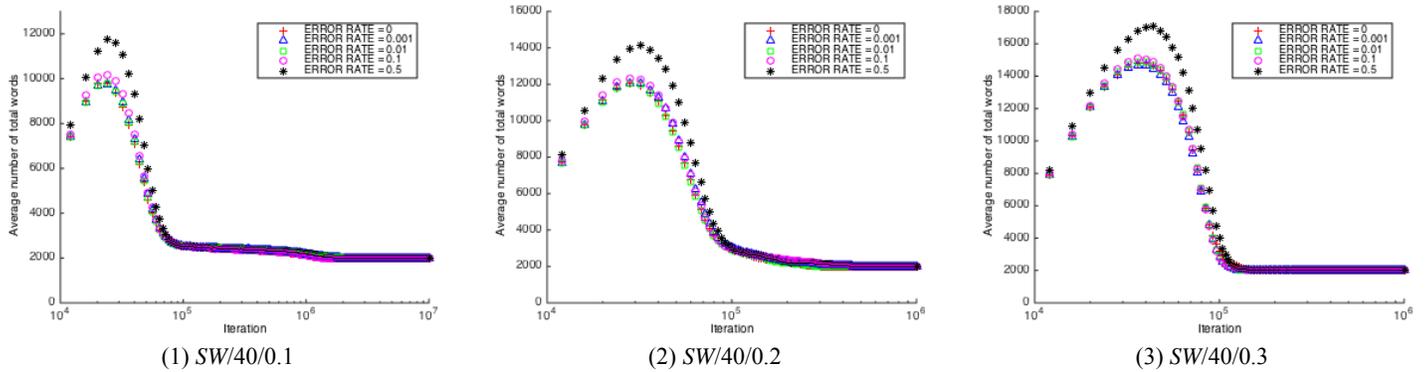

(1) $SW/40/0.1$    (2) $SW/40/0.2$    (3) $SW/40/0.3$

Fig. S1    The convergence process in terms of the number of total words in small-world networks with the number of neighborhoods $K = 40$

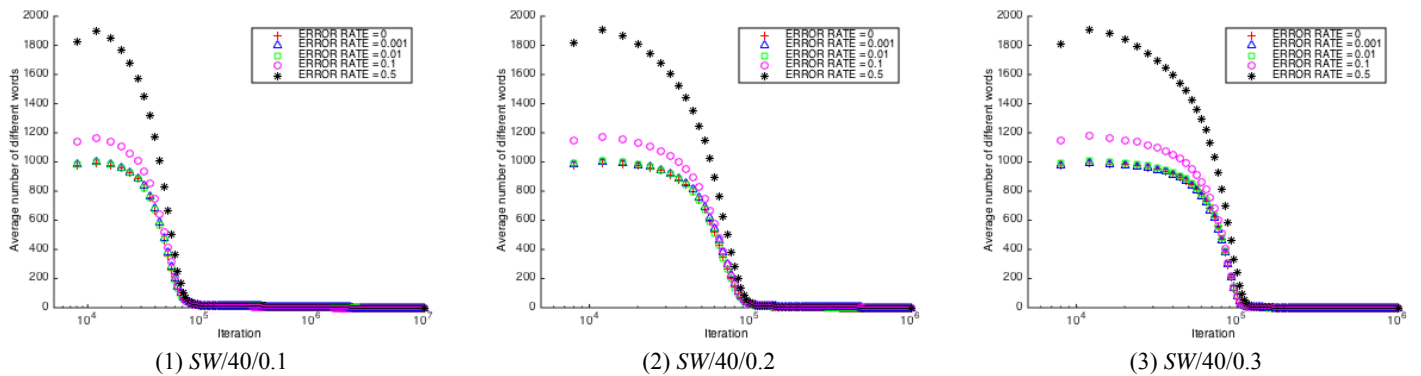

(1) $SW/40/0.1$    (2) $SW/40/0.2$    (3) $SW/40/0.3$

Fig. S2    The convergence process in terms of the number of different words in small-world networks with the number of neighborhoods $K = 40$

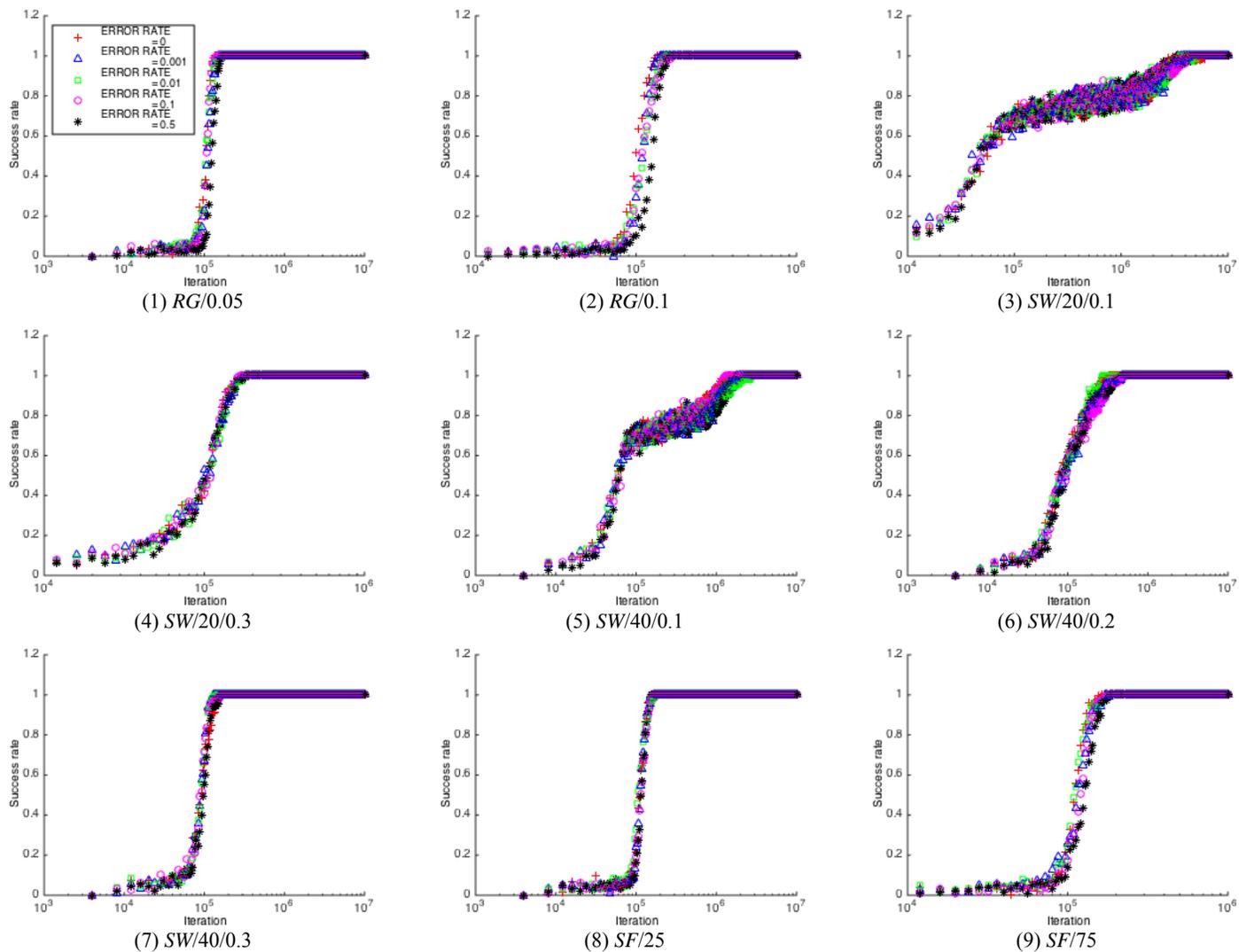

Fig. S3  Success rate curves including two random-graphs RG/0.05 and RG/0.1, five small-worlds SW/20/0.1, SW/20/0.3 and SW/40/{0.1, 0.2, 0.3}, and two scale-frees SF/25 and SF/75

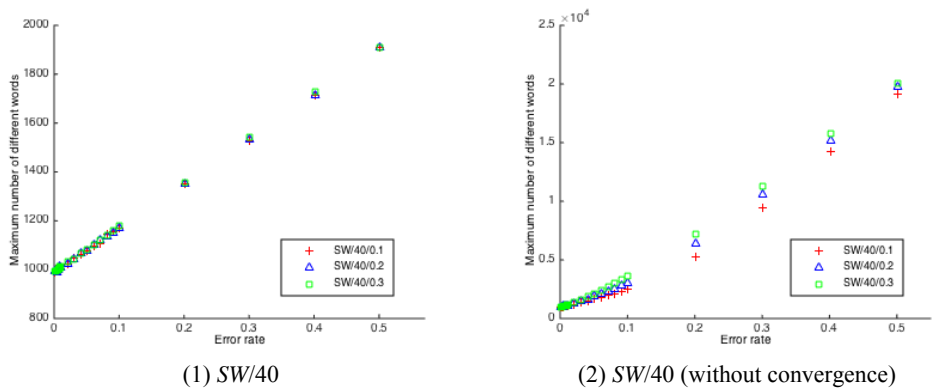

(1) SW/40     (2) SW/40 (without convergence)

Fig. S4  The relationship between the error rate and the maximum number of different words, in the topology of small-world with the number of neighborhoods K = 40

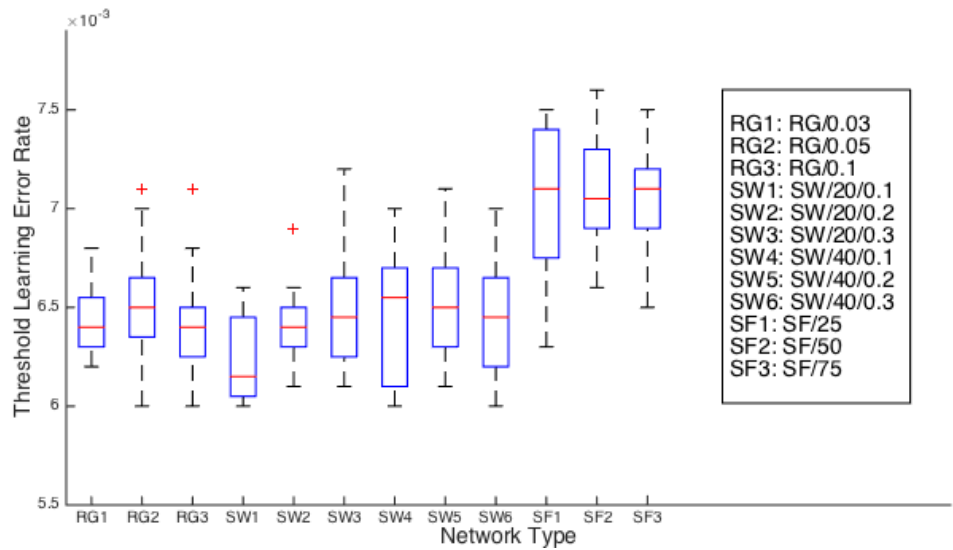

Fig. S5　The box plot of learning error rate threshold in different all twelve types of networks in this study

Table S2　The increment relationship between convergence time and different values of the error rate for small-word networks with number of neighborhoods $K = 40$ with different rewiring probabilities

| Net-works Error Rate | SW/40/0.1 | SW/40/0.2 | SW/40/0.3 |
|---|---|---|---|
| 0 | 1.0000 | 1.0000 | 1.0000 |
| 0.001 | +0.1785 | +0.1560 | −0.0376 |
| 0.002 | +0.1827 | +0.0905 | +0.0004 |
| 0.003 | +0.2084 | +0.1946 | −0.0433 |
| 0.004 | +0.1518 | +0.1762 | −0.0077 |
| 0.005 | +0.1642 | +0.1546 | −0.0250 |
| 0.006 | +0.1485 | +0.0026 | −0.0189 |
| 0.007 | +0.1749 | +0.1534 | −0.0548 |
| 0.008 | +0.0888 | +0.2110 | −0.0370 |
| 0.009 | +0.1443 | **−0.0570** | −0.0636 |
| 0.01 | +0.2334 | −0.0381 | −0.0516 |
| 0.02 | +0.1067 | +0.0641 | −0.0073 |
| 0.03 | +0.0574 | +0.1501 | −0.0510 |
| 0.04 | +0.0925 | **+0.2737** | **+0.0284** |
| 0.05 | +0.1268 | +0.1548 | −0.0384 |
| 0.06 | +0.1356 | +0.0809 | **−0.0868** |
| 0.07 | +0.1527 | +0.2576 | −0.0611 |
| 0.08 | **+0.2858** | +0.1552 | 0.0145 |
| 0.09 | +0.1217 | +0.0554 | −0.0054 |
| 0.1 | **−0.0002** | +0.2330 | −0.0312 |
| 0.2 | +0.1560 | +0.1912 | −0.0125 |
| 0.3 | +0.1110 | +0.2616 | −0.0060 |
| 0.4 | +0.0558 | −0.0328 | −0.0194 |
| 0.5 | +0.2607 | +0.1551 | +0.0209 |
| Summary | 22+/1− | 20+/3− | 4+/19− |

Table S3　Statistical division of the numbers of different incremental values in Table S2 into different intervals

| Increment value intervals | (−∞, −0.2) | [−0.2, −0.1) | [−0.1, 0) | [0, 0.1) | [0.1, 0.2) | [0.2, +∞) |
|---|---|---|---|---|---|---|
| Number of data | 0 | 0 | **23** | **13** | **24** | 9 |